# The impact of energy constraints on the medium access

Lazaros Gkatzikis   Georgios Paschos   Iordanis Koutsopoulos

**Abstract**

Contemporary mobile devices are battery powered and due to their shrinking size and increasing complexity operate on a tight energy budget. Thus, energy consumption is becoming one of the major concerns regarding the current and upcoming wireless communication systems. On the other hand, the available bandwidth resources are limited and modern applications are throughput demanding, leading thus to strong competition for the medium. In this direction, we consider a stochastic contention based medium access scheme, where the devices may choose to turn off for some time in order to save energy. We perform an analysis for a slotted ALOHA scenario and we show that the energy constraints, if properly exploited, may reduce contention for the medium. Our results give valuable insights on the energy–throughput tradeoff for any contention based system.

## I. INTRODUCTION

As mobile communications become part of our everyday life, new challenges for the system designers come to the foreground. First of all, the scarcity of bandwidth resources leads to extreme competition for the medium. Besides, the total energy dissipation by communication devices has been shown to amount to a significant portion of a nation's power profile, motivating efforts of per device energy economy. In an attempt to minimize their energy footprint and/or maximize the battery lifetime, existing wireless devices support radio sleep modes.

A generic wireless terminal consists of several circuit building blocks with the RF transceiver (radio) contributing significantly to the overall energy consumption. The RF transceiver itself consists of four subblocks. The transmit block that is responsible for modulation and up-conversion (i.e. transforms the baseband signal to RF), the receive block dedicated to the down-conversion and demodulation, the local oscillator that generates the required carrier frequency, and the power amplifier that amplifies the signal for transmission. Existing wireless devices support radio sleep modes that turn off specific subblocks, to minimize their energy consumption while inactive. For example, as shown in Table I the CC2420 transceiver ([4]) provides three different low power modes. In the deepest sleep mode, both the oscillator and the voltage regulator are turned off, providing hence the lowest current draw. However, this comes at the cost of the highest switching energy cost and the longest switching latency. On the other hand, the idle mode provides a quick and energy inexpensive transition back to the active state, but at the cost of higher current draw and consequently higher current consumption.

To address this tradeoff, the authors of [6] propose a scheme to dynamically adjust the power mode according to the traffic conditions in the network. They show that in a low traffic scenario, a deep sleep mode should be preferred since most of the time the nodes tend to be inactive. In a high traffic setting though, a "lighter" sleep mode is preferable, because frequent mode transitions incur high delay and energy costs, exceeding any energy saving coming from the low current draw.

In this direction, several energy aware MAC protocols have been proposed, either centralized or distributed ones, to resolve contention. However, most of them rely on the willingness of the nodes to comply with the protocol rules. Hence, they are vulnerable to selfish users that may deviate from the protocol in order to improve their own performance. Game theory comes as the ideal tool to model interactions among self–interested entities competing for common resources [7] and it has also been considered recently for medium access.

In [5], the authors study the Nash Equilibrium Points (NEP) in a slotted ALOHA system of selfish nodes with specific quality-of-service requirements. It has been observed that usually selfish behavior in medium access leads to suboptimal performance. For example, a prisoners dilemma phenomenon arises among selfish nodes using the generalized slotted Aloha protocols of [8]. A decrease in system throughput, especially when the workload increases due to the selfish behavior of nodes, is observed in [1], [3]. In an attempt to mitigate the effects of selfishness, the authors of [9] study the problem of minimizing the energy consumption for given throughput demands for a contention MAC. They show that whenever the demands are feasible, there exist exactly two Nash equilibrium points and derive a greedy mechanism that always converges to the best one.

TABLE I
SWITCHING TIME AND ENERGY CONSUMPTION OF A CC2420 RADIO

| Power mode | Switching time(ms) | Switching Energy ($\mu$J) | Current Consumption ($\mu$A) |
|---|---|---|---|
| Tx | 0 | 0 | 10000 |
| idle | 0.1 | 1.035 | 426 |
| power down | 1.2 | 42.3 | 40 |
| deep sleep | 2.4 | 85.7 | 0.02 |

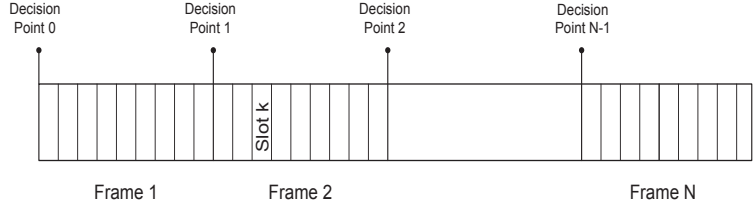

Fig. 1. The structure of a superframe

In this paper we introduce an additional level of decision making capturing the ON-OFF strategy of the terminals over the classic ALOHA game. Thus, we model contention for the medium as a game, where users with specific energy constraints *select both the proportion of time that they sleep and their medium access probabilities*. In this point we should mention that the main characteristics of the slotted ALOHA are also apparent in most contemporary contention-based systems, such as the 802.11. For example, all these systems exhibit a certain amount of inherent inefficiency. The throughput breaks down significantly, as the number of users and the message burstiness increase. In theory, an ideal CSMA/CA protocol should provide the same throughput, independent of the number of radios operating in a frame. In practice though, in order to avoid collisions some communication overhead has to be added, making hence the throughput a decreasing function of the number of operating radios. For instance, 802.11b achieves a 2 to 4Mbps effective throughput with only a few stations talking versus its maximum raw data rate of 5.5Mbps. A very accurate approximation of the actual throughput was derived in [2]. Consequently, our results can also be insightful for other contention based systems.

To the best of our knowledge, this is the first work that addresses the interplay between contention and energy consumption for systems that support sleep modes. The contributions of this paper can be summarized in the following:

- We quantify the interplay between contention and energy saving.
- We characterize the throughput optimal strategy, given the energy constraints.
- We develop a distributed approach that captures the notion of proportional to the energy budget fairness.
- We also formulate contention as a non-cooperative game among self–interested entities.
- We show that the resulting game has a unique NEP.
- We show that the energy constraints cause bounded Price of Anarchy.
- Based on the rationality of the users, we derive an improved alternative strategy and show that it has multiple NEPs.

## II. SYSTEM MODEL

We consider a communication scenario, where a set $\mathcal{N}$ of mobile terminals, with $|\mathcal{N}| = N$, wishes to transmit to a common destination (e.g. uplink to a Base Station). Time is slotted and within each timeslot each user may select either to transmit or to stay silent. Thus, the number of *active* users within the timeslot $t$ can be modeled as a stochastic process $N_a(t)$. We assume that medium access is performed probabilistically, according to a slotted ALOHA protocol, where a collision occurs whenever two or more terminals transmit concurrently. Each terminal has always buffered packets for transmission (i.e. saturated queue), but limited energy resources. Each device $i$ is characterized by an energy budget $\tilde{e}_i$, representing either its available battery power or the maximum energy it is willing to pay for. In order to save energy it can turn into a sleep mode, where most of the circuits are turned off. For analytical tractability we assume that each terminal may be in one out of two possible states, either ON or OFF.

In general, a mode transition incurs significant energy and time (delay) costs. Besides, due to hardware limitations the time required for a mode transition is of at least order of msec, much larger than the duration of a timeslot. Consequently, the mode transition at the timeslot level is neither feasible nor desirable and hence we introduce a new timescale, which we call frame, and where the mode switching takes place. Several timeslots constitute a frame and an arbitrary number of frames, say $N$, forms a superframe. The duration of the frame is dictated by the time required to obtain convergence to the ALOHA mean behaviour, whereas the horizon of operation can be arbitrarily large. As depicted in Fig. 1 the beginning of each frame is a decision point, where a node may change its operation mode. Inside any given frame, the nodes keep their mode fixed (either ON or OFF). Then, within the frame any active node may access the medium randomly according to a probability p. This probability is also assumed fixed on a per frame basis.

The control of a user $k$ for each superframe $j$ can be represented by a binary vector $\boldsymbol{q}_k(j) = \{0,1\}^N$, denoting the ON–OFF states, and an access probability $p_k$, i.e. we assume here that a user selects his access probability only once per superframe. In practice though, the mobile terminals are capable of making some decisions, but very rarely can exchange enough real time information to perform deterministic control and achieve the optimal behavior. A convincing example for using simple probabilistic controls is the framework of the 802.11 networks, prevailing in the world of computers, where the access behavior is dictated by probabilistic control. Similarly, by design choice we study a probabilistic version of the aforementioned problem which can be stated as follows.

Each user $i$ is characterized by a probability of being ON, denoted with $q_i$ and a medium access probability $p_i$. In matrix notation the strategy space can be written as $\mathcal{I} = \{\boldsymbol{p}, \boldsymbol{q}\}$, with $\boldsymbol{p} = [p_1, p_2, \ldots p_N]$ and $\boldsymbol{q} = [q_1, q_2, \ldots q_N]$. In our setting, the throughput of a user $i$ can be defined as the number of successfully exploited slots per unit time, and is a random variable with a mean of:

$$\bar{T}_i(p_i, q_i) = p_i q_i \prod_{j \in \mathcal{N}\setminus i}(1 - p_j q_j) \stackrel{\text{if } p_i q_i \neq 1}{=} \frac{p_i q_i}{1 - p_i q_i} \prod_{j \in \mathcal{N}}(1 - p_j q_j) \tag{1}$$

Note that the per user throughput is an increasing function of the decision variables $p_i$ and $q_i$, but decreasing in the number of terminals $N$ contending for the medium. The latter is in compliance with the classic ALOHA, but in practice also holds for CSMA/CA protocols.

The energy cost of user $i$ is a random variable with a mean value of $\bar{E}_i(p_i, q_i) = q_i(c_1 + c_2 p_i)$, where $c_1$ is the energy consumption of the ON state and $c_2$ the additional cost imposed by the transmission. Obviously, in order to be able to transmit, the node has to be in the ON state. Here, we have not considered the actual energy consumption of the transition itself.

## III. THE IMPACT OF CONSTRAINED ENERGY RESOURCES ON THE SYSTEM THROUGHPUT

In this general setting, we would like to find the ON–OFF and the medium access probabilities that maximize the collision–free utilization of the medium, and consequently the throughput of the system. This can be formally expressed as the following optimization problem:

$$\begin{aligned}
\underset{\mathcal{I}=\{\boldsymbol{p},\boldsymbol{q}\}}{\text{maximize}} \quad & \sum_{i=1}^{N} \bar{T}_i(p_i, q_i) \\
\text{s.t.} \quad & \bar{E}_i(p_i, q_i) \leq \tilde{e}_i \quad \forall i \\
& \{p_i, q_i\} \in [0, 1]^2 \quad \forall i
\end{aligned} \tag{2}$$

Throughout the paper, and without loss of generality, we assume that the users are ordered in decreasing energy budget, i.e. $\tilde{e}_1 \geq \tilde{e}_2 \geq \ldots \geq \tilde{e}_N$.

### A. Throughput optimal scheduling in energy constrained ALOHA with sleep modes

In the classic ALOHA setting, where no energy constraints exist, the throughput optimal strategy would be the one that eliminates contention. Thus, if we could force only a single user, say user $k$, to access the medium with probability $p_k = 1$ in each frame, we would achieve the maximum total throughput. In our scenario though, due to the energy constraints, the users may not be able to stay continuously ON (i.e $q_k = 1$) or to transmit with $p_k = 1$. Then, what is the best way to exploit the available energy resources? For each user we need to find the portion of energy to spend for staying ON during the frames and the portion used for transmitting within an ON frame.

*Lemma 1:* Out of all the throughput optimal strategies the most energy efficient ones are of the form $\mathcal{I}^* = \{\boldsymbol{1}, \boldsymbol{a}\}$ with $\boldsymbol{1} = [1, 1, \ldots 1]$ and $\boldsymbol{a} \in [0, 1]^N$. Thus, without loss of optimality we may restrict the strategy search space only to strategies where the nodes transmit continuously inside any ON frame.

*Proof:* Let $\hat{\mathcal{I}}$ be a feasible throughput optimal strategy and $a_i = \hat{p}_i \hat{q}_i$. From eq. 1 we may see that throughput depends only on the $pq$ products. Thus, the strategy $\mathcal{I}^* = \{\boldsymbol{1}, \boldsymbol{a}\}$ with $\boldsymbol{a} = [\hat{p}_1 \hat{q}_1, \hat{p}_2 \hat{q}_2, \ldots \hat{p}_i \hat{q}_i]$ achieves the optimal throughput, i.e. $\bar{T}(\mathcal{I}^*) = \bar{T}(\hat{\mathcal{I}})$.

Then, we prove that $\mathcal{I}^*$ is also feasible, as the most energy efficient strategy.

Any other feasible throughput optimal strategy $\hat{\mathcal{I}}$ can be written as an expression of $\boldsymbol{a}$ as $\{\hat{p}_i = \frac{a_i}{a_i + \delta_i}, \hat{q}_i = a_i + \delta_i\}$, with $0 < \delta_i \leq \min\{1 - a_i, \frac{\tilde{e}_i - a_i(c_1 + c_2)}{c_1}\}$. Thus, regarding the energy efficiency we have:

$$\begin{aligned}
\bar{E}\left(\hat{\mathcal{I}}\right) &= \sum_{i=1}^{N}(a_i + \delta_i)\left(c_1 + c_2 \frac{a_i}{a_i + \delta_i}\right) \\
&= \sum_{i=1}^{N} \delta_i c_1 + a_i(c_1 + c_2) = \bar{E}(\mathcal{I}^*) + c_1 \sum_{i=1}^{N} \delta_i \\
&> \bar{E}(\mathcal{I}^*).
\end{aligned}$$

This completes our proof. ∎

Based on Lemma 1, the optimization problem described by eq. 2 can be simplified to an expression that depends only on $\boldsymbol{q}$ leading to the objective function $\bar{T}(\boldsymbol{q}) = \sum_{i=1}^{N} q_i \prod_{j \in \mathcal{N} \setminus i} (1 - q_j)$ and constraints $0 \leq q_i \leq \tilde{q}_i = \min\left\{\frac{\tilde{e}_i}{c_1+c_2}, 1\right\}$; this can be further simplified into a problem of binary integer programming.

*Lemma 2:* The optimal solution is of the form $\boldsymbol{q}^* = \boldsymbol{b}^* \text{diag}[\tilde{q}_1, \tilde{q}_2, \ldots \tilde{q}_N]$ where $\boldsymbol{b}^*$ is a binary row vector.

*Proof:* The partial derivative of the objective function is given by:

$$\begin{aligned}
\frac{\partial \bar{T}}{\partial q_k} &= \prod_{j \in \mathcal{N} \setminus k} (1 - q_j) - \sum_{j \in \mathcal{N} \setminus k} q_j \prod_{l \in \mathcal{N} \setminus \{k,j\}} (1 - q_l) \\
&= \prod_{j \in \mathcal{N} \setminus k} (1 - q_j) - \sum_{j \in \mathcal{N} \setminus k} \frac{q_j}{1 - q_j} \prod_{l \in \mathcal{N} \setminus k} (1 - q_l) \\
&= \prod_{j \in \mathcal{N} \setminus k} (1 - q_j) \left(1 - \sum_{j \in \mathcal{N} \setminus k} \frac{q_j}{1 - q_j}\right)
\end{aligned}$$

We have thus shown that the sign of the partial derivative depends only in the parameter $\sum_{j \in \mathcal{N} \setminus k} \frac{q_j}{1 - q_j}$. This leads to the following decision making:

$$q_k = \begin{cases} \tilde{q}_k, & \text{if } \sum_{j \in \mathcal{N} \setminus k} \frac{q_j}{1 - q_j} < 1, \\ 0, & \text{otherwise.} \end{cases} \qquad (3)$$

■

*Lemma 3:* The optimal solution $\boldsymbol{b}^*$ is of the form $\boldsymbol{b}^* = [1, 1, \ldots 1, 0, 0, \ldots 0]$.

*Proof:* Assume that $\tilde{\boldsymbol{b}}$ of $k$ ones mixed with zeros is the throughput optimal binary vector. We can construct a new vector $\hat{\boldsymbol{b}} \in [0, 1]^N$, which has activated only the first $k$ users and gives identical throughput. To construct such a vector we need to move any isolated one, say from position $l$ of the initial vector to a zero position say $m$ with $m < l$, by setting $\hat{b}_m = q_l$. Then based on Lemma 2 we may fully activate or deactivate any of these users, getting thus better throughput. If this process is repeated some times, we finally end up with a vector of higher throughput. This leads us to a contradiction. ■

Based on the aforementioned lemmas we may derive the centralized Algorithm 1 that yields the throughput optimal probabilistic strategy and is of linear, in the number of users $N$, complexity. The main idea behind this algorithm is that contention may or may not be beneficial, depending on the energy constraints of the users. Namely, an additional user is useful if and only if the energy resources of the already active users are not sufficiently large, leaving thus the medium underutilized. An additional user introduces a gain due to the exploitation of the empty frames, but also a loss, due to the collisions whenever he is concurrently active within a frame with someone else. If the average gain is greater than the induced loss, it is beneficial for the system to be enabled.

---

**Algorithm 1** Optimal probabilistic frame scheduling

---
1: Order users in decreasing $\tilde{e}_i$. Without loss of generality,
   we reassign the indices such that $\tilde{q}_1 \geq \tilde{q}_2 \geq \ldots \geq \tilde{q}_N$
2: $\boldsymbol{q} \leftarrow \boldsymbol{0}$
3: $j \leftarrow 1$
4: **while** $j \leq N$ and $\sum_{i=1}^{j} \frac{q_i}{1 - q_i} < 1$ **do**
5: $\quad q_j \leftarrow \tilde{q}_j$
6: $\quad j \leftarrow j + 1$
7: **end while**

---

*Theorem 1:* Algorithm 1 yields the throughput optimal probabilistic strategy.

*Proof:* The optimality of 1 comes directly from Lemmas 1, 2 and 3.

■

Although, this approach maximizes the total throughput, it also introduces coordination and fairness issues; in particular, leads to a deterministic frame scheduling mechanism requiring extensive coordination among the users. Besides, it causes extremely unfair treatment of the users with low energy budget. Thus, it would not be easily applied to dynamic distributed environments as the ones considered here.

*B. A distributed fair algorithm*

As an answer to the aforementioned problems and in an attempt to capture the notion of proportional fairness we may substitute the original objective function with the following: $U(\boldsymbol{p}, \boldsymbol{q}) = \sum_{i=1}^{N} w_i \log \bar{T}_i$. The multiplicative factor $w_i$ can be used to balance the throughput among the users of the system at will. For example, the value $w_i = \frac{\tilde{e}_i}{\sum_{k \in \mathcal{N}} \tilde{e}_k}$ would allow us to split the throughput proportionally to the energy budget of the users. By proper reformulation, the objective function can be rewritten as $U(\boldsymbol{p}, \boldsymbol{q}) = \sum_{i=1}^{N} \log\left[(p_i q_i)^{w_i}(1 - p_i q_i)^{w_{-i}}\right]$, where $w_{-i} = \sum_{k \in \mathcal{N} \setminus i} w_k = 1 - w_i$. This is a separable per user function that leads to a fully distributed implementation, requiring minimal or even no information exchange. Actually the only information required is the value of the total energy available in the terminals, namely $\sum_{k \in \mathcal{N}} \tilde{e}_k$, which can also be estimated by each user through sensing. The solution of this optimization problem, in accordance to Lemma 1, is of the form $\mathcal{I}^* = \{\boldsymbol{1}, \boldsymbol{a}\}$ with $\boldsymbol{a} = [\min\{w_1, \tilde{q}_1\}, \min\{w_2, \tilde{q}_2\}, \ldots, \min\{w_N, \tilde{q}_N\}]$.

*C. A modified strategy*

Up to here we have assumed that each user makes a decision once for his strategy and applies it forever. As a result, user $k$ whenever active, transmits with $p_k = 1$, independently of the number of active users within a frame. Thus, whenever two or more users select to transmit within a frame they receive zero payoff, but consume energy. Based on these, a rational player would be expected to backoff whenever a collision is detected. Although, the terminal is not allowed to switch off in a crowded frame, due to the switching time overhead incurred, it may reduce its access probability. This way it would avoid spending energy on useless collisions and could utilize these savings for pursuing further contention–free frames. Building on this idea we propose the following modified strategy.

Any active user attempts a transmission within the first timeslot of the current frame. If the transmission succeeds he uses a medium access probability of $p_i = 1$. Otherwise he adjusts his strategy, and reduces his transmission probability to $\tilde{p}_i$. It can be shown that this new strategy always yields better throughput than the original one. The expressions for the throughput and the energy consumption are now respectively given by:

$$\bar{T}_i = q_i \left\{ (1 - \tilde{p}_i) \prod_{j \in \mathcal{N} \setminus i} (1 - q_j) + \tilde{p}_i \prod_{j \in \mathcal{N} \setminus i} (1 - \tilde{p}_j q_j) \right\} \quad (4)$$

$$\bar{E}_i = q_i \left\{ c_1 + c_2 \left[ \tilde{p}_i + (1 - \tilde{p}_i) \prod_{j \in \mathcal{N} \setminus i} (1 - q_j) \right] \right\} \quad (5)$$

Algorithm 2 yields the throughput optimal strategy, by categorizing the users of the system into three groups, namely the aggressive, the conservative and the passive ones. The former capture the medium whenever they are active (ON), the second transmit only whenever they sense an empty frame and the last do not participate at all. It can be shown that in the optimal scheduling there exists at least one aggressive and one conservative user. This algorithm is of exponential complexity and will be used only for comparison purposes.

## IV. GAME THEORETIC APPROACH

In the previous section we derived probabilistic MAC protocols that require coordination of the actions of the users involved. However, in an autonomous setting as the one considered here, it is not necessary that the individuals will comply with the rules imposed by the protocol. Actually, at least some of the users may exhibit selfish behaviour and select the strategy that maximizes their own utility, namely their individual throughput, at the expense of others. Thus, in this section we model our initial problem as a non-cooperative game.

A non-cooperative game is defined by a set of players, a set of strategies and a metric that indicates the preferences of the players over the set of strategies. In our case we have:

- **Players**: the N users
- **Strategies**: user's $i$ set of feasible medium access and ON–OFF probabilities $\mathcal{I}_i = \{p_i, q_i : \bar{E}_i(p_i, q_i) \leq \tilde{e}_i \text{ and } 0 \leq p_i, q_i \leq 1\}$
- **User preferences**: represented by a utility function $U_i(\mathcal{I}_i)$; peer $i$ prefers strategy $\hat{\mathcal{I}}_i$ to $\mathcal{I}_i$ iff $U_i(\hat{\mathcal{I}}_i) > U_i(\mathcal{I}_i)$.

**Algorithm 2** Modified optimal probabilistic frame scheduling

1: Search over $\mathcal{B} = \{\mathcal{A}, \mathcal{C}, \mathcal{P}\}$, i.e. the set of all the possible partitions of $\mathcal{N}$ of size 3, with $|\mathcal{A}| \geq 1$ and $|\mathcal{C}| \geq 1$ for the throughput optimal assignment:
2: **for all** $i \in \mathcal{A}$ **do**
3: $\quad \{\tilde{p}_i, q_i\} = \{1, \min\{\frac{\tilde{e}_i}{c_1+c_2}, 1\}\}$  % *aggressive ones*
4: **end for**
5: **for all** $k \in \mathcal{C}$ **do**
6: $\quad \{\tilde{p}_k, q_k\} = \{0, \min\{\frac{\tilde{e}_i}{c_1+c_2 \prod_{j \in \mathcal{N}\setminus k}(1-q_j)}, 1\}\}$ % *conservative*
7: **end for**
8: **for all** $j \in \mathcal{P}$ **do**
9: $\quad \{\tilde{p}_j, q_j\} = \{0, 0\}$  %*passive*
10: **end for**

---

For the initial optimization problem (eq. 2) the utility function of user $i$ is defined as $U_i(\mathcal{I}_i) = \bar{T}_i(p_i, q_i) = p_i q_i \prod_{j \in \mathcal{N}\setminus i}(1 - p_j q_j)$. By using the KKT conditions we may derive the following lemma:

*Lemma 4:* The throughput optimal strategy for user $i$ is $\{p_i, q_i\} = \left\{1, \min\{\frac{\tilde{e}_i}{c_1+c_2}, 1\}\right\}$. The resulting game has a unique Nash Equilibrium Point, described by the strategy $\mathcal{I}^* = \{\mathbf{1}, \boldsymbol{q}^*\}$, with $q_i^* = \frac{\tilde{e}_i}{c_1+c_2}$.

*Proof:* Since the utility is an increasing function of both $p_i$ and $q_i$ the energy constraint should be satisfied with equality at the optimum. Then, it can be shown through the KKT conditions that the optimal strategy is given by $\{p_i, q_i\} = \left\{1, \min\{\frac{\tilde{e}_i}{c_1+c_2}, 1\}\right\}$.

Given that the strategy of a user is independent of the actions of the other users and depends only on his own energy constraint deriving the resulting NEP is straightforward. ∎

This can be also justified by the fact that the throughput of user $i$ is an increasing function of $q_i$. Consequently, each individual will select its $q_i$ so as to satisfy the constraint with equality. On the other hand, since $T_i$ is an increasing function of $p_i$, it would select a transmission probability equal to 1. Thus, we may deduce that at the Nash equilibrium point we receive throughput, only when a single user is ON in a frame. Given the NEP of the game we may quantify the performance loss arising due to the selfishness of the individuals, by using so called *Price of Anarchy* (PoA) metric. This is the ratio of the value of the objective function at the global optimum to its value at the NEP and in our setting is given by:

$$\text{PoA} = \frac{\sum_{i \in \mathcal{S}} \frac{\tilde{e}_i}{c_1+c_2} \prod_{j \in \mathcal{S}\setminus i}(1 - \frac{\tilde{e}_j}{c_1+c_2})}{\sum_{i \in \mathcal{N}} \frac{\tilde{e}_i}{c_1+c_2} \prod_{j \in \mathcal{N}\setminus i}(1 - \frac{\tilde{e}_j}{c_1+c_2})} \geq 1, \qquad (6)$$

where $\mathcal{S}$ is the set of enabled users at the global optimum.

Whereas in the classic ALOHA games the PoA is unbounded, in our energy constrained ALOHA setting the PoA is generally bounded, since the energy constraints are imposing a fictitious pricing scheme. The only case where the PoA grows unbounded is whenever two or more users have infinite power. These two users will involuntarily act as jammers for each other and for all the others yielding hence zero system throughput.

### A. The modified strategy as a non-cooperative game of perfect information

Here, we consider the game arising from the modified strategy. In this new setting, we derive a non–cooperative game of perfect information, where in each iteration a user selects the best response to the others strategies. By *best response* we mean that each peer updates his decision variables, so as to maximize its utility function, as a response to the others' actions.

*Theorem 2:* The best response strategy of user $i$ is given by:

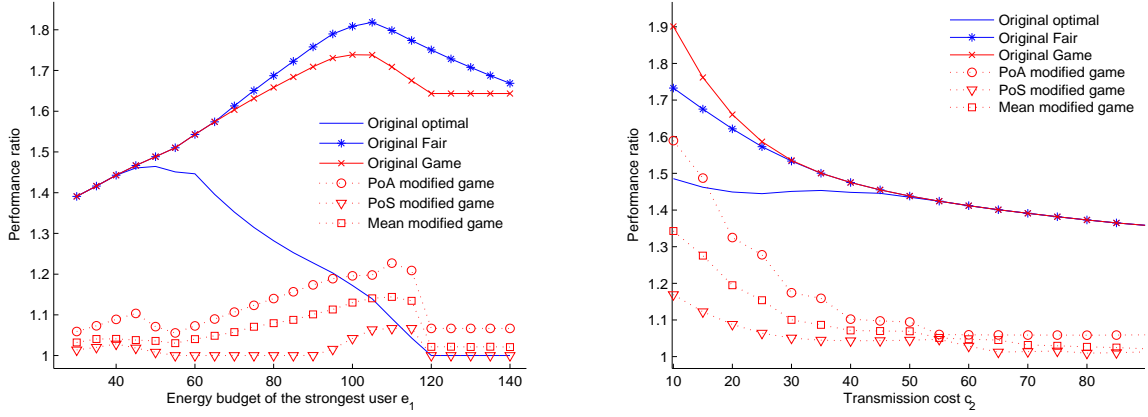

Fig. 2. The throughput degradation compared to the optimal

$$\tilde{p}_i = \begin{cases} 1, & \text{if } \left[c_1 + c_2 \prod_{j \in \mathcal{N} \setminus i}(1-q_j)\right] \prod_{j \in \mathcal{N} \setminus i}(1-\tilde{p}_j q_j) > (c_1+c_2) \prod_{j \in \mathcal{N} \setminus i}(1-q_j), \\ 0, & \text{otherwise.} \end{cases} \quad (7)$$

$$q_i = \frac{\tilde{e}_i}{c_1 + c_2 \left[\tilde{p}_i + (1-\tilde{p}_i) \prod_{j \in \mathcal{N} \setminus i}(1-q_j)\right]} \quad (8)$$

The arising modified game has multiple NEPs.

*Proof:* Since the utility is an increasing function of $\tilde{p}_i$ and $q_i$, the constraint needs to be satisfied with equality. Thus, we may replace $q_i$ from eq. 5 into the throughput expression, namely eq. 4. Then, the partial derivative of the objective function is given by the following expression:

$$\begin{aligned}\frac{\partial \bar{T}}{\partial \tilde{p}_i} &= \tilde{e}_i \frac{\left[\prod_{j \in \mathcal{N} \setminus i}(1-\tilde{p}_j q_j) - \prod_{j \in \mathcal{N} \setminus i}(1-q_j)\right]\left[c_1 + c_2 \prod_{j \in \mathcal{N} \setminus i}(1-q_j)\right] - c_2 \prod_{j \in \mathcal{N} \setminus i}(1-q_j)\left[1 - \prod_{j \in \mathcal{N} \setminus i}(1-q_j)\right]}{\left\{c_1 + c_2 \left[\tilde{p}_i + (1-\tilde{p}_i)\prod_{j \in \mathcal{N} \setminus i}(1-q_j)\right]\right\}^2} \\ &= \tilde{e}_i \frac{\left[c_1 + c_2 \prod_{j \in \mathcal{N} \setminus i}(1-q_j)\right]\prod_{j \in \mathcal{N} \setminus i}(1-\tilde{p}_j q_j) - (c_1+c_2)\prod_{j \in \mathcal{N} \setminus i}(1-q_j)}{\left\{c_1 + c_2\left[\tilde{p}_i + (1-\tilde{p}_i)\prod_{j \in \mathcal{N} \setminus i}(1-q_j)\right]\right\}^2}\end{aligned}$$

The sign of this expression depends only on $\prod_{j \in \mathcal{N} \setminus i}(1-\tilde{p}_j q_j)$ and $\prod_{j \in \mathcal{N} \setminus i}(1-q_j)$. As a result given the actions of the others, the objective function is either a strictly increasing or a strictly decreasing function of $\tilde{p}_i$. Thus, the best response strategy of user $i$ is given by 9. ∎

## V. NUMERICAL RESULTS

Here, we perform some simulations to quantify the throughput performance of the proposed schemes. We assume a network of $N = 5$ terminals with energy constraints given by $\tilde{e} = [30, 25, 15, 10, 5]$ and $\{c_1, c_2\} = \{50, 70\}$ units. We slightly abuse the definition of PoA, by using the modified optimal as the performance benchmark. Thus, the figures depict the performance degradation in comparison to the modified optimal. For the modified game we depict the PoA, the price of stability (PoS), defined as the throughput ratio of the optimum to the best NEP and the mean performance. Regarding the initial setting, we depict the performance degradation of the optimal, the fair and the initial game theoretic scheme.

Initially, we consider how the energy constraint of the most powerful user affects the performance of the system as a whole. As shown in the figure on the left, the additional power budget increases the performance degradation due to the additional collisions caused. The system stabilizes for $\tilde{e}_1 = c_1 + c_2$, where user 1 has sufficient energy to capture the whole medium on his own. On the right one, we depict the impact of the transmission cost $c_2$ on the performance. Here, we see that in a scenario of low energy constraints the increased transmission cost makes the users less aggressive, leading thus to reduced collisions. Both figures indicate that the modified strategy of backing off when a collision is detected may dramatically improve the performance.

## VI. Conclusion

This work is a first step towards characterizing the energy–delay tradeoff for mobile devices that support sleep modes and operate according to contention medium access schemes. We showed that energy constraints indirectly coordinate the actions of the players, and thus may reduce contention and lead to better exploitation of the medium. Here, we have assumed non–cooperative games of perfect information. However, there are contemporary wireless systems, where each involved entity has only a subjective belief on its opponents' strategies. The impact of incomplete information in our setting is an interesting topic of future study.

## References

[1] Eitan Altman, Rachid El Azouzi, and Tania Jiménez. Slotted aloha as a game with partial information. *Comput. Netw.*, 45(6):701–713, 2004.
[2] G. Bianchi. Performance analysis of the ieee 802.11 distributed coordination function. *Selected Areas in Communications, IEEE Journal on*, 18(3):535 –547, mar. 2000.
[3] R. El-Azouzi, T. Jiménez, E. S. Sabir, S. Benarfa, and E. H. Bouyakhf. Cooperative and non-cooperative control for slotted aloha with random power level selections algorithms. In *ValueTools '07: Proceedings of the 2nd international conference on Performance evaluation methodologies and tools*, pages 1–10, ICST, Brussels, Belgium, Belgium, 2007. ICST (Institute for Computer Sciences, Social-Informatics and Telecommunications Engineering).
[4] Texas Instruments. Texas instruments cc2420 radio transceiver. http://focus.ti.com/docs/prod/folders/print/cc2420.html.
[5] Youngmi Jin and G. Kesidis. Equilibria of a noncooperative game for heterogeneous users of an aloha network. *Communications Letters, IEEE*, 6(7):282 –284, jul. 2002.
[6] R. Jurdak, A.G. Ruzzelli, and G.M.P. O'Hare. Radio Sleep Mode Optimization in Wireless Sensor Networks. *IEEE Transactions on Mobile Computing*, 2010.
[7] I. Koutsopoulos, L. Tassiulas, and L. Gkatzikis. Client and server games in peer-to-peer networks. In *Quality of Service, 2009. IWQoS. 17th International Workshop on*, pages 1 –9, july 2009.
[8] R.T.B. Ma, V. Misra, and D. Rubenstein. An analysis of generalized slotted-aloha protocols. *Networking, IEEE/ACM Transactions on*, 17(3):936 –949, jun. 2009.
[9] I. Menache and N. Shimkin. Efficient rate-constrained nash equilibrium in collision channels with state information. In *INFOCOM 2008. The 27th Conference on Computer Communications. IEEE*, pages 403 –411, april 2008.